\documentclass[12pt]{article}
\usepackage{graphics}
\begin{document}
\begin{center}
{The Self-energy of Nucleon in the Pseudovector Coupling Pion-nucleon System
for the Electromagnetic Form Factor of Proton}
\end{center}
\begin{center}
{Susumu Kinpara}
\end{center}
\begin{center}
{\it National Institute of Radiological Sciences \\ Chiba 263-8555, Japan}
\end{center}
\begin{abstract}
The electromagnetic form factor of proton is studied following our previous calculation of the anomalous magnetic moment.
The non-perturbative term is incorporated to correct the $Q^2$ dependence. 
\end{abstract}
\vspace{4.mm}
\section*{\normalsize{1 \quad Introduction}}
\hspace*{4.mm}
The meson-exchange model for the relativistic nuclear many-body system is one of the subjects
which attracts our interest.
To obtain the results in the framework of the quantum fields the interaction between nucleons plays a decisive role. 
Particularly the lightest meson the pion has the longest range 
and the degree of freedom is indispensable to calculate the observables on the two-body system.
\\\hspace{4.mm}
In the case of the finite density such as the finite nuclei 
a lot of experimental data suggest the pseudovector coupling interaction to understand the properties
among two possible types of the interactions.
Since the form of the wave packet self-consistently determined suppresses the high momentum transfer component 
the derivative on the pion field does not bring in the difficulty of the divergences 
taking into account the correlation between two nucleons.
\\\hspace{4.mm}
Without the effect of the cut-off created under the nuclear medium
it is necessary to remove the divergences for the perturbative expansion in free space.
The prescription of the counter term has not been accomplished for the pseudovector coupling of the pion-nucleon system.
Then it is useful to examine the relation between the quantities constructed by the operators of the fields
irrespective of the strength of the interaction.
An interesting character of the non-perturbative relation is the appearance of the non-perturbative term
inherent in the breaking of the conservation of the axial-vector current.
\\\hspace{4.mm}
The application of the non-perturbative term to the self-energy is significant
and suitable for correcting the internal off-shell state of the nucleon propagator.
The non-perturbative term in the self-energy generates another one of the divergences.
It cancels out the divergence in the perturbative part which is
not removed completely by the counter terms of the mass and the wave function.
The resulting finite quantity is expected to supply corrections 
along with the ones by the method of the perturbative expansion. 
\\\hspace{4.mm}
The self-energy has an effect on the electromagnetic properties of nucleon explained by the vertex function.
They are connected by the non-perturbative relation 
analogous to the Ward-Takahashi (W-T) relation in the quantum electrodynamics.
In addition to the Dirac part of the form factor supposing the point-like structure of proton 
the terms of the self-energy contribute only to the anomalous part.
The sign and the magnitude of the result by the lowest-order approximation to the self-energy are found to be appropriate 
to understand the value of the anomalous magnetic moment of nucleon 
in conjunction with the corrections by the perturbative expansion for the vertex.
On the other hand the present formulation does not give the momentum transfer dependence adequately.
It is our purpose of the present study to investigate the momentum dependence of the electromagnetic form factor
incorporating the non-perturbative term.
\\
\section*{\normalsize{2 \quad The electric and the magnetic form factors}}
\hspace*{4.mm}
In our previous study the electromagnetic vertex function of nucleon has been calculated in the lowest-order approximation
by using the currents of pion and nucleon \cite{Kinpara}.
At the limit of the four-momentum transfer of nucleon $Q^2 \rightarrow 0$ 
the anomalous part gives the anomalous magnatic moment of nucleon.  
The normalization of the Dirac part is automatically satisfied by the W-T relation such as $F_1(0) = 1$
between the on-shell states.
\\\hspace{4.mm}
The property $F_1(Q^2) \equiv 1$ remains unchanged by including the self-energy of the nucleon propagator 
which the W-T relation contains
since the term is only for the anomalous part $F_2(Q^2)$ 
and furthermore it does not depend on the $Q^2$ as shown in the appendix.
The use of the lowest-order term in the series of $\gamma\cdot k - M$ for the self-energy 
provides the constant value as $F_2(Q^2) = 2 + O(m_\pi^2/M^2)$ for proton. 
Taking account of it and admitting to adjust the pion-nucleon coupling constant $f_\pi$ 
the correction by the perturbative expansion makes possible to reproduce the value of the magnetic moment.
The decrease in $f_\pi$ from $f_\pi = 1$ required could be related to the perturbative higher-order corrections 
for the pion-nucleon vertex part. 
\\\hspace{4.mm}
The $Q^2$ dependence of the anomalous part $F_2(Q^2)$ is evaluated by using the expressions in ref. \cite{Kinpara}.
The region of the space-like photon ($Q^2 < 0$) suffices to study the electromagnetic properties such as the extended structure of proton for the electron-proton elastic scattering \cite{Rosenbluth}.
As well as the $Q^2 = 0$ case 
the $x$- and $y$- integral of the analytic function is performed without the appearance of the divergences.
\\\hspace{4.mm}
In order to compare with the experiment the form factors $F_1(Q^2)$ and $F_2(Q^2)$ are substituted with
the electric and the magnetic form factors $G_E(Q^2)$ and $G_M(Q^2)$ which are defined by
\begin{eqnarray}
\qquad G_E(Q^2) \equiv F_1(Q^2) + \frac{Q^2}{4 M^2}F_2(Q^2)
\end{eqnarray}
\begin{eqnarray}
G_M(Q^2) \equiv F_1(Q^2) + F_2(Q^2).
\end{eqnarray}
The $F_2(Q^2)$ contains the anomalous magnetic moment $\kappa$ and normalized to be $F_2(0) = \kappa$ in the present study.
At present the values of our calculation 
are $\kappa_p = 1.745$ for proton and $\kappa_n = -3.003$ for neutron \cite{Kinpara} 
in units of the nuclear magneton using the coupling constant of the pseudovector pion-nucleon interaction $f_\pi =1.0$. 
We need to lower the value of $f_\pi$ to reproduce the experimental values of the proton and neutron simultaneously.
It is owing to the isospin dependence and the coefficients of the pion and the nucleon current parts
which are roughly same in size and have different signs.
\\\hspace{4.mm}
The dependence of $G_E(Q^2)$ on $-Q^2$ is shown in Fig. 1. The relation $F_1(Q^2) \equiv 1$ is used in Eq. (1)
by virtue of the W-T relation.
The result of the calculation is compared with the experimental data 
which are represented by the dipole form factor $G_D(Q^2) = (1-Q^2/0.71)^{-2}$ \cite{Gayou}.
The theoretical curve predicted by the calculation of the lowest-order perturbative expansion
decreases more slowly than that of $G_D(Q^2)$.
Accordingly the radius of the charge of proton $\sqrt{<r^2>}$ is too small which is given by the relation
$\sqrt{6 \, d G_E(Q^2)/d Q^2 \vert_{Q^2=0}}= 0.34 \, (0.81) \, {\rm fm} $.
The value in parentheses is the one taken from the $G_D(Q^2)$.
\\\hspace{4.mm}
To derive the size of proton in terms of the meson-exchange model 
the cloud of pions surrounding the point-like structure of proton is to be improved by some effects furthermore.
Since the correction of the $\sim Q^2$ term in $F_2(Q^2)$ changes only the $\sim (Q^2)^2$ term for $G_E(Q^2)$ 
it does not improve the $\sqrt{<r^2>}$ value.
Then the $F_1(Q^2)$ is required to have the $Q^2$ dependence making us explain the properties of $G_D(Q^2)$.
While the shift of $F_1(Q^2)$ from $F_1(Q^2) \equiv 1$ is assumed to be large as seen from Fig. 1, 
it is not sensitive to the ratio $(1+\kappa_p) G_E(Q^2)/G_M(Q^2)$ 
because of the cancellation of the additional terms.
On the other hand the change of the $F_2(Q^2)$ by adding a $\sim Q^2$ term inevitably makes the ratio leave from
the form $(1+\kappa_p) G_E(Q^2)/G_M(Q^2) \sim 1-0.196\,(-Q^2-0.345)$
obtained by the fit of the theoretical curve a in Fig. 1 within the range $0<-Q^2<2$.
\\
\section*{\normalsize{3 \quad Inclusion of the non-perturbative term}}
\hspace*{4.mm}
The electromagnetic vertex of nucleon is extended so as to include the non-perturbative term.
The pion-nucleon-nucleon three-point vertex is modified as
\begin{eqnarray}
\Gamma_i(p,q) = \Gamma_i(p,q)_{per} + \Gamma_i(p,q)_n
\end{eqnarray}
\begin{eqnarray}
\Gamma_i(p,q)_n = G(p)^{-1}\,\tau_i\,\gamma_5 +\tau_i\,\gamma_5\,G(q)^{-1}
\end{eqnarray}
where $G(p)$ and $G(q)$ are the nucleon propagators with the outgoing ($p$) and the incoming ($q$) momenta respectively. 
The part $\Gamma_i(p,q)_{per}$ is determined by the perturbative expansion and
the term of the lowest-order is given by 
$\Gamma_i(p,q)_{per} = \tau_i\,\gamma_5\,\gamma\cdot(p-q) + O((f_\pi/m_\pi)^2)$.
One of the features of the vertex is that the joint use of $\Gamma_i(p,q)_{per}$ and $\Gamma_i(p,q)_{n}$ 
is identical to that of the pseudoscalar coupling in the lowest-order 
when the self-energy of the propagator is neglected in Eq. (4).
\\\hspace{4.mm}
The non-perturbative term is applied to the electromagnetic form factor of proton
by the replacement 
$\Gamma_i(p,q)_{per} \rightarrow \Gamma_i(p,q)$
of the three-point vertex.
In the process mediated by the pion current the part what is modified is divided into four pieces as  
\begin{eqnarray}
\Gamma_i(p,k)\,G(k)\,\Gamma_j(k,q)\,=\, \tau_i\tau_j \,(A+B+C+D)
\end{eqnarray} 
\begin{eqnarray}  
A \equiv \gamma_5\,\gamma\cdot(p-k)\,G(k)\,\gamma_5\,\gamma\cdot(k-q)
\end{eqnarray} 
\begin{eqnarray}  
B \equiv \gamma_5\,\gamma\cdot(p-k)\,G(k)\,(G(k)^{-1}\,\gamma_5+\gamma_5\,G(q)^{-1})
\end{eqnarray} 
\begin{eqnarray}  
C \equiv (G(p)^{-1}\,\gamma_5+\gamma_5\,G(k)^{-1})\,G(k)\,\gamma_5\,\gamma\cdot(k-q)
\end{eqnarray} 
\begin{eqnarray}  
D \equiv (G(p)^{-1}\,\gamma_5+\gamma_5\,G(k)^{-1})\,G(k)\,(G(k)^{-1}\,\gamma_5+\gamma_5\,G(q)^{-1}).
\end{eqnarray} 
When the non-perturbative term is dropped the parts $B$, $C$ and $D$ vanish 
and the form factor is reduced to the previous one obtained by the procedure of the perturbative expansion. 
The former part $A+B$ corresponds to the self-energy appearing in the W-T relation diagrammatically.
It has been dealed with to add the constant shift to the anomalous magnetic moment of proton 
and the part $A+B$ is not included in the calculation.
\\\hspace{4.mm}
Our interest is the part $C+D$ which is independent of the W-T relation.
It is expected to provide an additional $Q^2$ dependence of the $G_E(Q^2)$.
For the expression in Eq. (5) is put between the Dirac spinors $\bar{u}(p)$ and $u(q)$ the inverses of the propagators
$G^{-1}(p)$ and $G^{-1}(q)$ in Eqs. (6)$\,\sim\,$(9) are set equal to zero 
and also the $\gamma \cdot q$ is replaced with the nucleon mass $M$. 
Then $C+D$ results in 
\begin{eqnarray}
C+D \rightarrow -2\,M - \gamma_5 \, {\it\Sigma(k)} \, \gamma_5.
\end{eqnarray} 
The $\it\Sigma(k)$ is the self-energy of the nucleon propagator.
In Eq. (10) the relation $\gamma_5\,\gamma\cdot k = - \gamma\cdot k \, \gamma_5$ is used without
the extra term on the shift of the dimension of space-time. 
$\it\Sigma(k)$ is in a series of $\gamma\cdot k - M$ 
starting from the quadratic order $\sim (\gamma\cdot k - M)^2$ by virtue of the condition for the on-shell state of nucleon.
The series are added up and the $C+D$ becomes the closed form 
$C+D = \alpha(k^2)\,\gamma\cdot k+\beta(k^2)$ 
in terms of the coefficients $\alpha(k^2)$ and $\beta(k^2)$ as a function of $k^2$.
\\\hspace{4.mm}
The correction of the electromagnetic form factor of nucleon mediated by the pion current is given as 
\begin{eqnarray}
&&\Gamma_\pi^\mu (p,q) = -2\,i\,\tau_3\,(\frac{f_\pi}{m_\pi})^2 \int \frac{d^4 k}{(2 \pi)^4} \,(-2 k +p+q)^\mu 
\nonumber\\
&&\times\,(\alpha(k^2)\,\gamma\cdot k+\beta(k^2))\, {\it\Delta}^{0}(k-p)\, {\it\Delta}^{0}(q-k)
\end{eqnarray}
in which ${\it\Delta}^{0}$ is the free propagator of pion. 
The $f_\pi$ and $m_\pi$ are the coupling constant of the pseudovector type of the interaction and the pion mass respectively.
Since $\alpha(k^2)$ and $\beta(k^2)$ are not in the form of the polynomials they are substituted by constant values
mentioned below to apply the formula of the dimensional regularization method to the integral over $k$ in Eq. (11).
Then $\alpha(k^2)$ and $\beta(k^2)$ are expanded in the series of $k^2-M^2$
and only the lowest order is left as $\alpha(k^2) \approx \alpha(M^2)$ and $\beta(k^2) \approx \beta(M^2)$.
The region of the $k$-integral is required to be divided into two parts 
because the radius of convergence of the series is finite.
The integral of the outer region is dropped by spreading out the boundary to infinity.
\\\hspace{4.mm}
The approximate way to the $k$-integral is on the basis of the fact 
that the anomalous part of the form factor by the pion current contribution
agrees with each other between the pseudoscalar and the pseudovector coupling interactions
provided that the four-momentum of the internal nucleon is limited to the on-shell state ($k^2 = M^2$).
In actual the calculation of the anomalous magnetic moment shows the results of the two kinds of the interactions 
are nearly identical neglecting the extra term related to the gamma matrix $\gamma_5$ 
in the pseudovector coupling \cite{Kinpara}.
\\\hspace{4.mm}
Under the procedure mentioned above the $k$-integral is done by the formula 
and the result is expressed by using the variable of the integral $z$ as
\begin{eqnarray}
\Gamma_\pi^\mu (p,q) 
= - \tau_3 \, \gamma^\mu \, \frac{\alpha(M^2)\,m_\pi^2}{2\,M^2} \, (\frac{2Mf_\pi}{4\pi m_\pi})^2 \,
[ \, \eta \, (1-\frac{Q^2}{6m_\pi^2}) \nonumber\\
-\int_0^1 d z \, \{1-\frac{Q^2}{m_\pi^2}z(1-z)\}\,{\rm log}\{1-\frac{Q^2}{m_\pi^2}z(1-z)\} \, ]
\end{eqnarray}
\begin{eqnarray}
\eta \equiv \frac{2}{\epsilon} - \gamma + 1 - {\rm log} \frac{m_\pi^2}{4 \pi \mu^2}
\end{eqnarray}
\begin{eqnarray}
&&\int_0^1 d z \, \{1-\zeta\,z(1-z)\}\,{\rm log}\{1-\zeta\,z(1-z)\} \nonumber\\
&&=\frac{1}{18\,\sqrt{\zeta(4-\zeta)}}\{\,(-24+5\,\zeta)\sqrt{\zeta(4-\zeta)} 
+\,6\,(4-\zeta)^2\,{\rm arctan}\sqrt{\frac{\zeta}{4-\zeta}}\,\} \nonumber\\
&&=\,-\,\frac{\zeta}{6}+O(\zeta^2).
\end{eqnarray}
In Eq. (13) the $\gamma = 0.577\cdots$ is the Euler's constant. 
The two parameters $\epsilon$ and $\mu$ are generated by
the shift of the space-time dimension as $4 \rightarrow 4-\epsilon$ in the dimensional regularization method.
Different from the calculation of the anomalous magnetic moment in ref. \cite{Kinpara} 
there exists the divergence and which is not removed by the counter term.
It may be attributed to the approximation of $\alpha(k^2) \rightarrow \alpha(M^2)$ replacing with a constant value 
although it vanishes as $\alpha(k^2) \sim 1/k^2$ at $k^2 \rightarrow \infty$.
At the moment the $\eta$ is set to zero as $\eta \approx 0$ to proceed the calculation in the present study.
\\\hspace{4.mm}
The term relating to $\beta(M^2)$ does not contribute to the $\sim \gamma^\mu$ part in $\Gamma_\pi^\mu (p,q)$ 
as verified by carrying out the integral over $z$ after applying the formula of the $k$-integral.
The self-energy $\it\Sigma(k)$ is represented as follows
\begin{eqnarray}
{\it\Sigma(k)} =  c_1(k^2) -\gamma\cdot k \, c_2(k^2)
\end{eqnarray}
\begin{eqnarray}
c_1(k^2) \equiv \frac{M(k^2+M^2)(M^2+m_\pi^2-k^2/2)+M^3 k^2}{(M^2+m_\pi^2-k^2/2)^2-M^2 k^2 /4}
\end{eqnarray}
\begin{eqnarray}
c_2(k^2) \equiv \frac{M^2(k^2+M^2)/2+2M^2(M^2+m_\pi^2-k^2/2)}{(M^2+m_\pi^2-k^2/2)^2-M^2 k^2 /4}
\end{eqnarray}
and then $\alpha(M^2) = -\,c_2(M^2) = -2M^2/m_\pi^2$.
\\\hspace{4.mm}
Determining the value of $\alpha(M^2)$ the form factor of proton $F_1(Q^2)$ yields
\begin{eqnarray}
F_1(Q^2) = 1 + \frac{1}{6} \, (\frac{2Mf_\pi}{4\pi m_\pi})^2 \, \frac{Q^2}{m_\pi^2} +O((Q^2)^2).
\end{eqnarray} 
The change by $F_1(Q^2)$ is excessive and the curve of $G_E(Q^2)$ decreases faster than $G_D(Q^2)$ as seen in Fig. 1.
Consequently the charge radius of proton is $\sqrt{<r^2>} = 1.55\, {\rm fm} $ and it is about twice as large as
the experimental value.
The overestimate is expected from the result of the calculation in which the strength of the interaction is too large 
to reproduce the anomalous magnetic moment of neutron \cite{Kinpara}.
\\\hspace{4.mm}
The higher-order corrections for the three-point vertex may improve the result
by suppressing the value of the coupling constant $f_\pi$ effectively.
The approximation of $\alpha(k^2)$ to $\alpha(M^2)$ may be the cause of the gap 
in addition to the occurence of the divergence.
It has been found that the decrease of the variable $k^2$ in $\alpha(k^2)$ from $M^2$ only 9 $\%$ 
achieves to reproduce the experimental value of $\sqrt{<r^2>}$.
\\\hspace{4.mm}
The use of the lowest-order form ${\it\Sigma(k)} \approx M (M^2+m_\pi^2)^{-1} (\gamma\cdot k-M)^2$ is interesting
since it is appropriate to the calculation of the anomalous magnetic moment of proton. 
The correction of $F_1(Q^2)$ becomes small in comparison with the one by the exact form of the self-energy. 
It gives the value of the charge radius $\sqrt{<r^2>} = 0.41\, {\rm fm} $ and the $Q^2$ dependent curve
because of the change of $\alpha(M^2)$ from $-2M^2/m_\pi^2$ to $-2M^2/(M^2+m_\pi^2)$.
\\
\section*{\normalsize{4 \quad Concluding remarks }}
\hspace*{4.mm}
The degree of freedom of pion is necessary to understand the $Q^2$ dependence 
of the electromagnetic form factor determined by the elastic scattering of electron from proton.
In order to break the relation $F_1(Q^2) \equiv 1$ 
the pion-nucleon-nucleon vertex part has been extended to incorporate the non-perturbative term.
The excess of the root mean square radius indicates the need of some further effects such as the vertex corrections
or improvement of the $k$-integral.
Results of the calculation are influenced by the approximate form of the self-energy chosen 
and then the self-consistent equation is significant to investigate the role of pion accurately. 
Besides the $Q^2$ term it is essential to give the strong $(Q^2)^2$ dependence by some effect
so that the $G_E(Q^2)$ remains to be at the positive values in the intermediate energy region.
\\
\section*{\normalsize{ Appendix }}
\hspace{4.mm}
The photon-nucleon-nucleon three-point vertex part $\Gamma(p,q)$ is associated 
with the nucleon propagators $G(p)$ and $G(q)$ as
\\\begin{eqnarray}
(p-q)\cdot\Gamma(p,q) = \frac{1+\tau_3}{2}\,(G(p)^{-1}-G(q)^{-1}).
\end{eqnarray}
In the right-hand side the contribution of the self-energy is given by the series of 
\begin{eqnarray}
(\gamma \cdot p - M)^n - (\gamma \cdot q - M)^n \qquad (n \geq 2)
\end{eqnarray}
and it consists of two parts 
\begin{eqnarray}
= (p-q)\cdot \{ (p+q)\, A_n + \gamma \, B_n \},
\end{eqnarray}
\begin{eqnarray}
A_n \equiv \frac{(-M)^n}{p^2-q^2} \, \{ \frac{(1+\sqrt{p^2/M^2})^n+(1-\sqrt{p^2/M^2})^n}{2}- (p^2 \rightarrow q^2) \}
\nonumber\\
- \frac{\gamma\cdot(p+q)}{2 M} \cdot \frac{(-M)^n}{p^2-q^2} \, 
\{ \frac{(1+\sqrt{p^2/M^2})^n-(1-\sqrt{p^2/M^2})^n}{2 \sqrt{p^2/M^2}}- (p^2 \rightarrow q^2) \}
\end{eqnarray}
\begin{eqnarray}
B_n \equiv \frac{(-M)^{n-1}}{4} \, \{ \frac{(1+\sqrt{p^2/M^2})^n-(1-\sqrt{p^2/M^2})^n}{\sqrt{p^2/M^2}}
 + (p^2 \rightarrow q^2) \}.
\end{eqnarray}
When two momenta $p$ and $q$ of the outgoing and the incoming nucleons 
are at the on-shell ($\gamma\cdot p \,\rightarrow M$, $\gamma\cdot q \,\rightarrow M$), the coefficients $A_n$ and $B_n$
have their respective values as $A_n \rightarrow (-2 M)^{n-2}$ and $B_n \rightarrow (-2 M)^{n-1}$.
In Eq. (21) the quantity becomes $(p+q)^\mu\, A_n + \gamma^\mu \, B_n \rightarrow -(-2M)^{n-2} \, i \sigma^{\mu\nu}(p-q)_\nu$
by using the relation of the Gordon decomposition.
Then it contributes only to the anomalous part of the form factor.

\newpage
\begin{figure}
\begin{center}
\scalebox{0.5}{\includegraphics{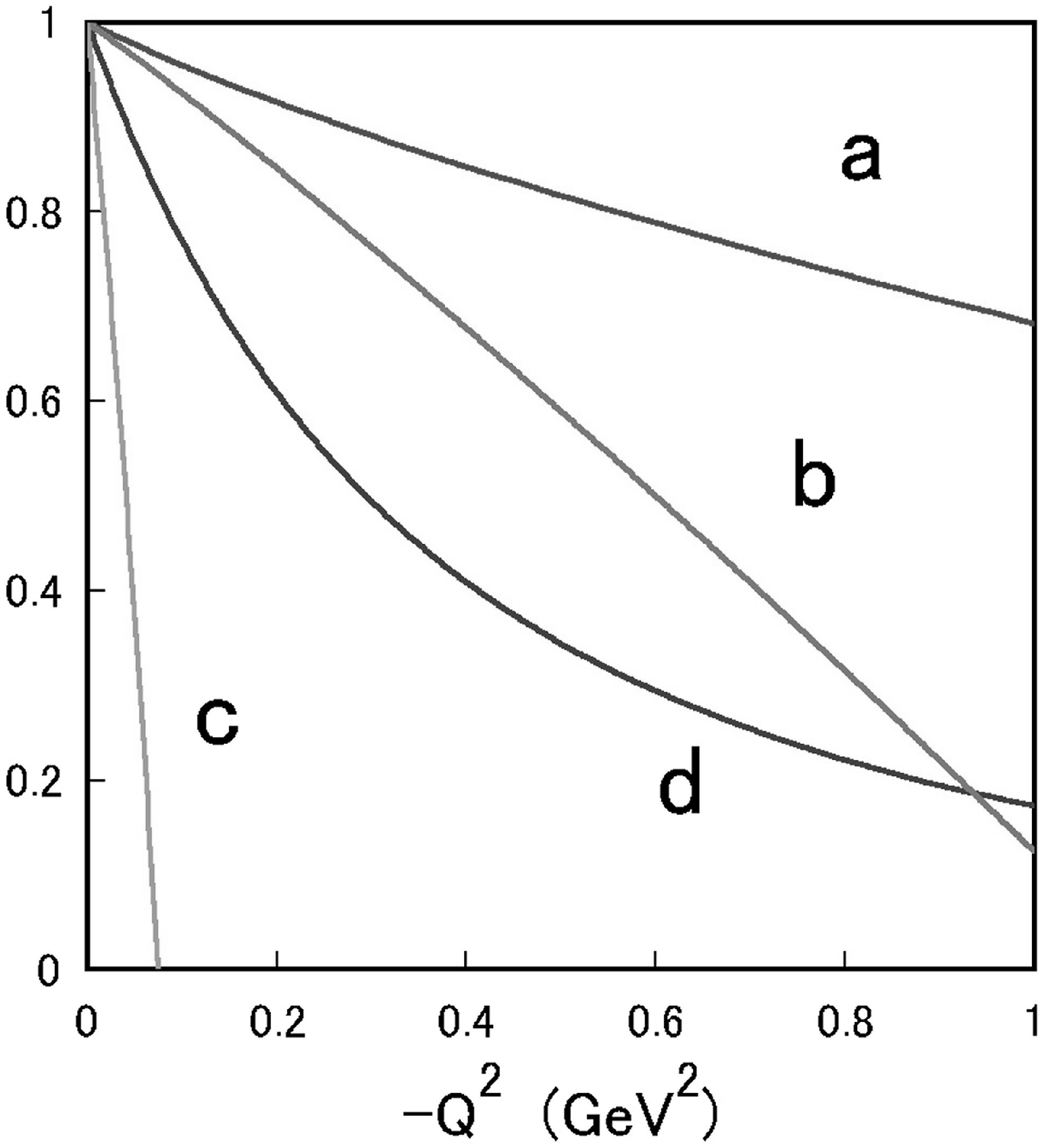}}
\caption{
The electric form factors $G_E(Q^2)$ are shown as a function of $-Q^2$. 
a:\,\,the result of the calculation with $F_1(Q^2) \equiv 1$. 
b:\,\,including the lowest-order self-energy effect on $F_1(Q^2)$.
c:\,\,including the full self-energy effect on $F_1(Q^2)$.
d:\,\,the dipole form factor $G_D(Q^2) = (1-Q^2/0.71)^{-2}$.
}
\end{center}
\end{figure}
\end{document}